\documentclass[journal]{IEEEtaes}
\usepackage{graphicx} 
\usepackage{amsthm}
\usepackage{amsfonts}
\usepackage{amssymb,bm}
\usepackage{mathrsfs}
\usepackage{mathtools}
\usepackage{amsmath}
\usepackage{subcaption}
\usepackage{balance}
\usepackage{url}
\usepackage{graphicx}
\usepackage{textcomp}
\usepackage{accents}
\usepackage{url}
\usepackage{gensymb}
\usepackage{xcolor}
\usepackage{algorithm}
\usepackage{algorithmic}

\begin{document}
\title{Low-Complexity Super-Resolution Signature Estimation of XL-MIMO FMCW Radar}
\author{Chandrashekhar Rai}
 \affil{Indian Institute of Technology Delhi,  India}
\author{Arpan Chattopadhyay}
 \affil{Indian Institute of Technology Delhi, India}
\authoraddress{
Both authors are with Electrical Engineering, Indian Institute of Technology (IIT) Delhi, India. Email: \emph{ird600941@ee.iitd.ac.in, arpanc@ee.iitd.ac.in}.\\
The work of Chandrashekhar Rai was supported by the Visvesvaraya Post Doctoral Fellowship scheme, Digital India Corporation, MeitY, India. The work of Arpan Chattopadhyay was supported in part by the Science and Engineering Research Board (SERB) India under Grant CRG/2022/003707, in part by Indo-French Centre for the Promotion of Advanced Research under Grant IFC/7150/2023, in part by SYSTRA MVA Consulting India Private Ltd., India under Project RP04860N, and in part by the Qualcomm Technologies, Inc, USA under Project FT/2024/11/37.}

\maketitle

\begin{abstract}
    Extremely Large-Scale (XL) multiple input multiple output (MIMO) antenna systems combined with ultra-wide signal bandwidth (BW) offer the potential for ultra-high-resolution sensing in frequency modulated continuous wave (FMCW) radars. However, the use of ultra-wide BW results in significant spatial delays across the array aperture, comparable to the range resolution, leading to the spatial wideband effect (SWE). SWE introduces coupling between the range and angle domains, rendering conventional narrowband signal processing techniques ineffective for target signature estimation. In this paper, we propose an efficient super-resolution signature estimation technique for XL-MIMO FMCW radars operating under SWE, leveraging compressive sensing (CS) methods. The proposed 2D CS-based approach offers low computational complexity, making it highly suitable for real-time applications in large-scale radar systems. Numerical simulation results validate the superior performance of the proposed method compared to existing wideband and narrowband estimation techniques.
\end{abstract}

\section{Introduction}
Modern radar systems, particularly in automotive applications and emerging sensing technologies, demand ultra-high resolution for reliable environment perception and object detection \cite{patole2017automotive}. This has given rise to a high level of interest in extra-large multiple-input multiple-output (XL-MIMO) radar systems, which leverage hundreds or thousands of antenna elements to achieve fine angular and range resolution while maintaining wide coverage \cite{wu2021intelligent}. Enhancing radar resolution necessitates deploying ultra-large arrays combined with high modulation bandwidths \cite{park2024spatial,han2023range,xi2021joint,durr2020range,hu2023range,wang2021direction,wu2024coffee,janoudi2023signal}. However, this combination introduces two major challenges: (i) the spatial wideband effect (SWE)  where the spatial delay across the radar aperture becomes comparable to the range resolution \cite{wang2018spatial}, and (ii) the near field effect (NFE)  where, at large array scales, the planar wavefront assumption fails for nearby targets \cite{wei2021channel}. While the NFE becomes increasingly important for extremely large arrays and short-range targets, its explicit modeling and joint treatment with SWE are beyond the scope of this work and will be investigated in future studies. On the other hand, the SWE significantly complicates joint range and angle-of-arrival (AoA) estimation since the range and angle terms of the intermediate frequency (IF) signal become cross-coupled, making conventional signal processing approaches ineffective. In narrowband MIMO-FMCW radars, joint range-angle estimation is commonly achieved using 2D frequency estimation or sequential 1D techniques with manageable complexity \cite{fang2019joint}. Conversely, under SWE in wideband radars, the decoupling assumption fails, and the radio scene's sparsity structure evolves into a block sparsity model \cite{wang2018spatial,park2024spatial,han2023range,hu2023range}.

Several classic wideband AoA estimation techniques exist, such as the coherent signal subspace method (CSSM) \cite{wang1985coherent}, interpolated array techniques \cite{friedlander1993direction}, and test of orthogonality of projected subspaces (TOPS) \cite{yoon2006tops}. Recent works like \cite{xi2021joint} adopt subspace-based separable approaches utilizing algorithms such as RELAX \cite{li1997angle}. While effective in narrowband or moderate-size array scenarios, these subspace methods rely on prior knowledge of the number of targets and impose computational burdens unsuited for the scale of XL-MIMO systems.

To the best of our knowledge, no existing work in the literature addresses low-complexity, joint range-angle estimation in XL-MIMO FMCW radars under SWE without assuming the number of targets. Motivated by this gap, we first develop a comprehensive system model capturing SWE effects in XL-MIMO FMCW radar systems, which takes into account the inherent range-angle coupling in far-field scenarios.  To address the associated high-dimensional estimation challenges, a scalable and computationally efficient algorithm is developed for the joint estimation of the number of targets, their ranges, AoAs, and complex reflection coefficients without requiring prior target count information. This is accomplished by extending the conventional two-dimensional orthogonal matching pursuit (2D-OMP) technique to exploit the block-sparse structure of spatial wideband radar scenes, enabling robust super-resolution signature estimation. Extensive simulation results are presented to validate the proposed method, demonstrating superior estimation accuracy and computational efficiency compared to existing subspace-based and wideband AoA estimation techniques across various XL-MIMO radar configurations.

\textbf{Notations:} Scalars are denoted by small letters (e.g., $x$), vectors by bold lowercase letters (e.g., $\mathbf{x}$), and matrices by bold uppercase letters (e.g., $\mathbf{X}$). The transpose and conjugate transpose of a matrix $\mathbf{X}$ are denoted by $\mathbf{X}^{\mathrm{T}}$, and $\mathbf{X}^{\mathrm{H}}$, respectively. The Euclidean norm of a vector is denoted by $\|\cdot\|_2$.

\section{System Model}
The linear frequency modulated (LFM) chirp transmitted by a transmit antenna is modeled as
\par\noindent\small
\begin{align}
s(t) = \exp{\left(j2\pi\left(f_{c}t+\frac{\gamma}{2}t^{2}\right)\right)}, \;\; 0\leq t\leq T_{ch},
\end{align}
\normalsize
where $f_c$ is the carrier frequency, $\gamma$ is the chirp slope, and $T_{ch}$ is the duration of a chirp.

We consider a target scene comprising $K$ far-field, non-fluctuating point targets, where the $k$-th target's range, radial velocity, and angle of arrival (AoA) are denoted by $R_{k}$, $\nu_{k}$, and $\theta_{k}$, respectively. In our framework, each receiver independently processes the signal received from every transmitted chirp, since time-division multiplexing (TDM) ensures orthogonality among the transmitted signals. Accordingly, we first consider the received signal component at the $m$-th receiver for a single chirp transmitted from the $n$-th transmitter, given by
\par\noindent\small
\begin{align}
r_{n,m}(t) = \sum_{k=1}^{K} a_{k} \, s(t-\tau^{k}_{n,m}), \;\; 0\leq t\leq T_{ch},
\end{align}
\normalsize
where $a_{k}$ is the complex amplitude proportional to the $k$-th target’s radar cross-section (RCS), and $\tau^{k}_{n,m}$ is the total time delay for the $k$-th target’s reflected signal.

\begin{figure}
    \centering
    \includegraphics[width=0.75\linewidth]{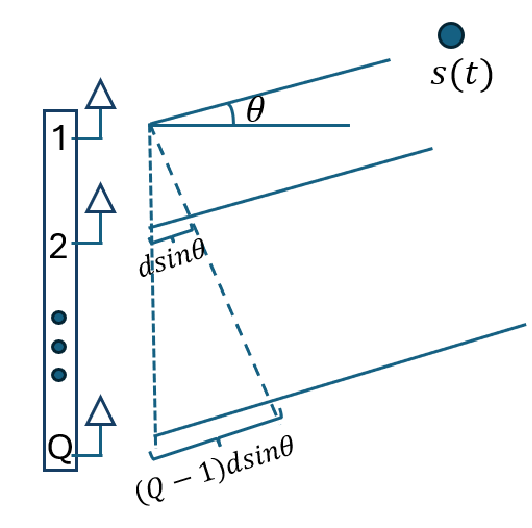}
    \caption{The ULA illustration of the MIMO radar}
    \label{fig:system_model}
\end{figure}

Considering the virtual array corresponding to the MIMO configuration as shown in Fig. \ref{fig:system_model}, the received signal at the $q$-th element of the virtual uniform linear array (ULA), with inter-element spacing $d$, can be written as
\par\noindent\small
\begin{align}
r_{q}(t) = \sum_{k=1}^{K} a_{k} \, s(t-\tau^{k}_{q}), \;\; 0\leq t\leq T_{ch},
\end{align}
\normalsize
where the total delay $\tau^{k}_{q}$ consists of a range delay, a Doppler-induced delay, and an angular delay, expressed as
\par\noindent\small
\begin{align}
\tau^{k}_{q} = \tau^{k}_{R} + \tau^{k}_{D} + \tau^{k}_{q,\theta}, \label{eqn:delay components}
\end{align}
\normalsize
with $\tau^{k}_{R}=2R_{k}/c$, $\tau^{k}_{D}=2\nu_{k}t/c$, and, under the far-field assumption, $\tau^{k}_{q,\theta}\approx qd\sin{(\theta_{k})}/c$.

At the $q$-th virtual receiver, the received signal $r_{q}(t)$ is mixed with the transmitted chirp $s(t)$ to obtain the IF signal as
\par\noindent\small
\begin{align}
    y_{q}(t)=s(t)r^{*}_{q}(t)=\sum_{k=1}^{K}a_{k}^{*}&\exp{\left(j2\pi\gamma\tau^{k}_{q}t\right)} \exp{\left(-j\pi\gamma(\tau^{k}_{q})^{2}\right)}
    \nonumber\\
    &\;\;\times \exp{\left(j2\pi f_{c}\tau^{k}_{q}\right)},
\end{align}
\normalsize
which is then sampled at sampling frequency $f_{s}$ to yield the (discrete) fast-time measurements
\par\noindent\small
\begin{align}
    y_{q}[n]=\sum_{k=1}^{K}a_{k}^{*}\underbrace{\exp{\left(j2\pi\gamma\tau^{k}_{q}\frac{n}{f_{s}}\right)}}_{\textrm{Term-I}}
    \underbrace{\exp{\left(-j\pi\gamma(\tau^{k}_{q})^{2}\right)}}_{\textrm{Term-II}}\underbrace{\exp{\left(j2\pi f_{c}\tau^{k}_{q}\right)}}_{\textrm{Term-III}}.\label{eqn:discrete IF signal}
\end{align}
\normalsize

We, henceforth, deal with only discrete-time signals and use $n$ to denote the discrete-time index with $0\leq n\leq N-1$ where $N=\lceil f_{s}T_{ch}\rceil$. Moreover, throughout this work, only stationary targets are considered, and as a result, the Doppler-induced delay is neglected. Since the Doppler-induced delay is affected by SWE, further investigations are required, which we leave for future research. Next, we examine terms I, II, and III.

\textit{Term-I:} We define $\Omega_{R}^{k}\doteq\gamma \tau^R_k/f_s$ and normalized angular/spatial frequency $\Omega^{k}_{\theta}\doteq d\sin(\theta_{k})/\lambda$, $BW = \alpha f_c$, $\gamma = BW/T_c$. 
\par\noindent\small
\begin{align} 
    \exp{\left(j2\pi\gamma\tau_q^k \frac{n}{f_s}\right)} = &\exp{\left(j2\pi\frac{\gamma}{cf_s}(2R_k+qd\sin(\theta_k))n\right)}\nonumber\\
    &= \exp{\left(j2\pi\Omega_R^kn\right)}\exp{\left(j2\pi\frac{\Omega^k_{\theta}}{f_c}\gamma\frac{qn}{f_s}\right)}\nonumber\\ &=\exp{\left(j2\pi\Omega_R^kn\right)}\exp{\left(j2\pi\Omega^k_{\theta}\frac{\alpha}{N}qn\right)}.
\end{align}
\normalsize

\textit{Term-II:} For practical radar system parameters \cite{lovescu2020fundamentals}, the quadratic term-II is usually ignored. Substituting \eqref{eqn:delay components} in Term-II we get
\par\noindent\small
\begin{align}
    \exp{(-j\pi\gamma(\tau_q^k)^2)} &= \exp{\left(-j\pi\gamma \frac{4R_k^2}{c^2}\right)}\exp{\left(-j\pi\gamma\frac{q^2d^2\sin^2\theta_k}{c^2}\right)}\nonumber\\
    &\exp{\left(-j2\pi\gamma\frac{2R_kqd\sin\theta_k}{c^2}\right)}.
\end{align}
\normalsize
Here, $\exp{(-j\pi\gamma {4R_k^2}/{c^2})}$ does not vary with antenna index and hence can be absorbed in $\widetilde{a}_{k}$. Further, the maximum value of the remaining two terms in the exponents is close to zero and hence can be ignored (This can be verified numerically, for example by letting $R_k = 20 m$, $\theta_k = 90^{\circ}$, $Q = 128$, and $\gamma = 100 MHz/\mu s$).

\textit{Term-III:} Substituting \eqref{eqn:delay components} in Term-III, we obtain
\par\noindent\small
\begin{align}
    \exp{\left(j2\pi f_{c}\tau^{k}_{q}\right)}=\exp{\left(j2\pi f_{c}\tau^{R}_{k}\right)}\exp{\left(j2\pi\frac{qd\sin(\theta_{k})}{\lambda}\right)}.
\end{align}
\normalsize
Here, $\exp{\left(j2\pi f_{c}\tau^{R}_{k}\right)}$ does not vary with antenna elements and hence, can be included in $\widetilde{a}_{k}$ in \eqref{eqn:2D_swb}. Hence, we can write the equivalent IF signal in \eqref{eqn:discrete IF signal} with spatial wideband assumption as
\par\noindent\small
\begin{align}
    y_{q}[n]\approx\sum_{k=1}^{K}\widetilde{a}_{k}&\exp{(j2\pi\Omega^{k}_{R}n)}\exp{(j2\pi\Omega^{k}_{\theta}q)}\nonumber\\
    &\;\;\times \underbrace{\exp{(j2\pi\frac{\alpha}{N}\Omega^{k}_{\theta}qn)}}_{SW\;Term}+w_{q}[n],\label{eqn:2D_swb}
\end{align}
\normalsize
where $\widetilde{a}_{k}=a_{k}^*exp(j\pi\gamma(\tau^{R}_{k})^{2})exp(-j2\pi f_{c}\tau^{R}_{k})$ and $w_q[n]$ is additive white Gaussian noise (AWGN) which follows $\mathcal{CN}(0,\sigma^2)$. The spatial wideband (SW) Term in \eqref{eqn:2D_swb} is of particular interest in this work. It should be noted that the SW term depends on three main parameters- (i) system BW $\alpha$, (ii) number of array elements $Q$, and (iii) angle of arrival $\theta$. Specifically, for large BW and array elements, the spatial delay across the array is either large or comparable to the range resolution and hence the coupled SWE has to be considered in the system. For smaller BWs, $\alpha \rightarrow 0$, the SW term is approximated by 1, and the equivalent MIMO-FMCW IF signal becomes a mixture of 2D complex exponential tones as

\begin{equation}
    y_{q}[n]=\sum_{k=1}^{K}\widetilde{a}_{k}\exp{(j2\pi\Omega^{k}_{R}t)}\exp{(j2\pi\Omega^{k}_{\theta}q)}+w_q[n].
    \label{eqn:2D_snb}
\end{equation}

Finally, we define $\lbrace K,\Tilde{a}_k,\Omega_R^k, \Omega_{\theta}^k \rbrace$ as the signature of the radio scene.
It is important to highlight that range-angle estimation for spatial narrowband radar systems can be performed directly through the 2D spectral estimation of \eqref{eqn:2D_snb} \cite{rai2025low}. An analogous problem for sparse linear arrays has been addressed in \cite{rai2025multi}, with theoretical guarantees on signal recovery. However, these traditional narrowband techniques cannot be directly applied to spatial wideband radar systems, due to the time-antenna index coupling introduced by the SW term in \eqref{eqn:2D_swb}.

\section{Proposed Methodology}

We propose a novel signature estimation algorithm specifically designed to address the SWE in XL-MIMO FMCW radar systems. The proposed approach enables the application of the conventional narrowband 2D-OMP algorithm in spatial wideband scenarios by compensating for the inherent time-antenna index coupling introduced by the SWE term. The algorithm operates in two stages: it first performs a DFT-based coarse signature estimation to obtain preliminary target parameter estimates, followed by a CS-based fine-tuning stage that refines these estimates for enhanced resolution and accuracy. This hybrid strategy effectively mitigates the limitations of narrowband techniques under spatial wideband conditions while preserving computational efficiency suitable for large-scale radar arrays.

\subsection{DFT-based coarse estimation}
We can show that the 2D-DFT of the spatial narrowband IF signal in \eqref{eqn:2D_snb} can be represented as a 2D-sparse matrix whereas the 2D-DFT of the spatial wideband IF signal in \eqref{eqn:2D_swb} can be represented as a 2D-block sparse matrix in the range-angle domain \cite{wang2018spatial}. 
The $(u,v)^{th}$ element of 2D-DFT of the IF signal in \eqref{eqn:2D_swb} is defined as 
\par\noindent\small
\begin{align}
    \mathbf{Y}|_{(u,v)}\doteq \sum_{k=0}^{K-1}\sum_{n=0}^{N-1} \sum_{q=0}^{Q-1} \Big(\widetilde{a}_{k}\exp{(j2\pi\Omega^{k}_{R}n)}\exp{(j2\pi\Omega^{k}_{\theta}q)}&
    &\nonumber\\\underbrace{\exp{(j2\pi\frac{\alpha}{N}\Omega_{k}^{\theta}qn)}}_{SW\;Term}+w_q[n]\Big)
    \; \times \exp{\left(-j\frac{2\pi}{Q}uq\right)} \exp{\left(-j\frac{2\pi}{N}vn\right)},
    \label{eqn:2D_transform}
\end{align}
\normalsize

where $u$ is the angle bin index and $v$ is the range bin index.
Due to the presence of the coupled SW term in \eqref{eqn:2D_transform}, the two series with respect to $n$ and $q$ cannot be separated, and a closed-form solution for range-angle estimation is not achievable. Instead, the range-angle response forms a block corresponding to each point target. The spread of each block depends on the target’s AoA, the system BW, and the number of antenna elements in the MIMO-FMCW radar.

The number of target-induced blocks in the range-angle response can be estimated using techniques such as clustering \cite{rai2025artificial}, support detection \cite{bevelander2025divide}, or custom peak-finding algorithms \cite{rai2025low}. In this work, we assume that no two target blocks overlap under the SWE condition, ensuring distinct separation for detection. To identify these blocks, we employ the 2D peak search method proposed in \cite{MATLABCode} to locate the peaks of $||\mathbf{Y}_k||_2^2$ within the range-angle response, thereby estimating the number of targets $\hat{K}$. The peak location for each target block provides a coarse estimate of its corresponding range bin ($\hat{u}_k$) and AoA bin ($\hat{v}_k$), in the neighborhood $\mathcal{N}_k$ corresponding to the $k$-th target’s return.

It should be noted, however, that due to the spatial wideband effect, these detected peak locations do not precisely correspond to the true coarse range and angle bins. The bin positions are affected by SWE-induced migration, which requires further coarse bin correction as suggested in \cite{rai2025two}. However, in this work, a separate coarse correction mechanism is not required that adds to the robustness of the proposed method. Consequently, the coarse estimates of the normalized range and spatial frequencies are obtained as $\hat{\Omega}_R^k = \hat{u}_k/N$ and $\hat{\Omega}_{\theta}^k = \hat{v}_k/Q$, respectively. The coarse signature estimate for the scene is thus represented by the set $\left\lbrace \hat{K}, \hat{\Omega}_R^k, \hat{\Omega}_{\theta}^k \right\rbrace$.

\vspace{-0.5 cm}
\subsection{CS-based fine-tuned estimation}
The fine-tuned parameters of each target are estimated separately in each iteration over the estimated number of targets. An initial approximation of the target’s signature is used to refine these estimates over successive iterations, progressively improving the accuracy of the parameters using the 2D-OMP method. Once this refinement is complete, the target is removed from the observation set before proceeding to the next iteration.

For this, we first initialize $\Check{y}_q[n] = y_q[n]$, then for the $k$-th target, we compensate for the SWE using the coarse angle estimate $\hat{\Omega}_{\theta}^k$ as
\par\noindent\small
\begin{align}
    \Tilde{y}_q[n] = &\Check{y}_q[n] \times\exp{\left(-j2\pi\frac{\alpha}{N}\hat{\Omega}^{k}_{\theta}qn\right)}\\
    \Tilde{y}_q[n]=\sum_{j=0}^{K-1}\widetilde{a}_{j}&\exp{(j2\pi\Omega^{j}_{R}n)}\exp{(j2\pi\Omega^{j}_{\theta}q)}\nonumber\\
    \times\exp{\left(j2\pi\frac{\alpha}{N}\Omega_{j}^{\theta}qn\right)}
    &\exp{\left(-j2\pi\frac{\alpha}{N}\hat{\Omega}_{\theta}^{k}qn\right)} + \Tilde{w}_q[n] \nonumber\\
     = \sum_{j=0}^{K-1}\widetilde{a}^{*}_{k}\exp{(j2\pi\Omega^{j}_{R}t)}&\exp{(j2\pi\Omega^{j}_{\theta}q)} \underbrace{\exp{\left(j2\pi\frac{\alpha}{N}(\Omega_{\theta}^j-\hat{\Omega}_{\theta}^{k})qn\right)}}_{Residue\; Term}\nonumber\\
     + \Tilde{w}_q[n]&.
     \label{eqn:conjugation}
\end{align}
\normalsize

For the close values of coarse angle estimates ($\Omega_{\theta}^{j} \approx \hat{\Omega}_{\theta}^{k}$), the residue term in \eqref{eqn:conjugation} becomes close to unity and hence the SWE is compensated for the $k$-th path. It should be noted here that the power for the $k$-th target is enhanced significantly compared to other targets due to near-perfect SWE compensation, and the IF signal corresponding to this target can be approximately  recovered by putting an appropriate threshold, i.e., update the $\Tilde{y}_q[n]$

\[
\Tilde{y}_q[n] \leftarrow 
\begin{cases}
\Tilde{y}_q[n], & \text{if } |\Tilde{y}_q[n]| \geq \gamma \\
0,   & \text{if } |\Tilde{y}_q[n]| < \gamma.
\end{cases}
\]

Equivalently, the IF signal for the $k$-th path, converted to the narrowband signal model, can be written as  
\begin{align}
\Tilde{y}^k_q[n] = \widetilde{a}_{k}\exp{(j2\pi\Omega^{k}_{R}n)}&\exp{(j2\pi\Omega^{k}_{\theta}q)}+\Tilde{w}_q[n].
\label{eqn:2D_swb_conjugated}
\end{align}

\begin{figure*}  
    \centering
    \begin{subfigure}[b]{0.31\linewidth}
        \centering
        \includegraphics[width=\linewidth]{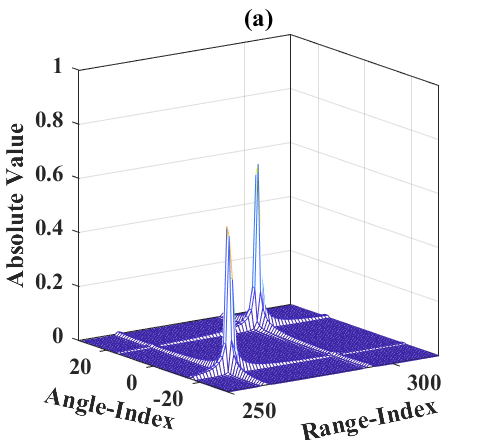} 
    \end{subfigure}
    \hfill
    \begin{subfigure}[b]{0.31\linewidth}
        \centering
        \includegraphics[width=\linewidth]{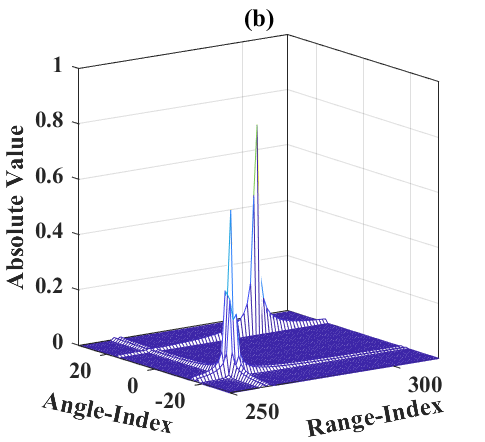}
    \end{subfigure}
    \hfill
    \begin{subfigure}[b]{0.31\linewidth}
        \centering
        \includegraphics[width=\linewidth]{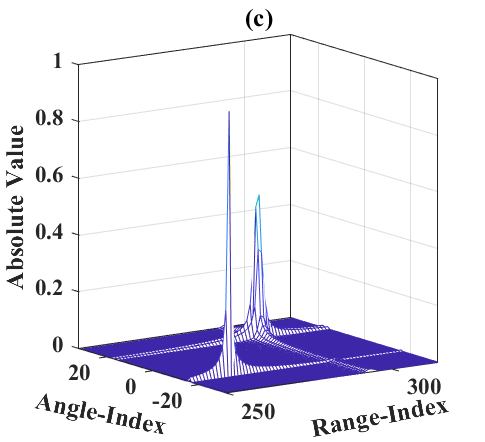}
    \end{subfigure}
    \caption{(a) Angle-Delay response and the SWE for the two paths (b) SWE compensation and power enhancement for Target 1 (c) SWE compensation and power enhancement for Target 2.}
    \label{fig:example1}
\end{figure*}

This process can be illustrated with an example, as shown in Fig.~\ref{fig:example1}. We consider two targets, where Target~1 is located at $[8.5\,\mathrm{m},\, 30^{\degree}]$ and Target~2 at $[9.5\,\mathrm{m},\, 60^{\degree}]$. The system bandwidth is set to $0.05 f_c$, with $f_c = 77$\,GHz, and the number of fast-time and spatial samples is $N = 512, ~Q = 64$. The two unit-amplitude targets produce spread responses in the angle-delay domain due to the SWE, as depicted in Fig.~\ref{fig:example1}(a). The coarse angle estimates for Target~1 and Target~2 are $30.5^{\degree}$ and $61.045^{\degree}$, respectively. Spatial wideband compensation, performed using \eqref{eqn:conjugation}, is shown in Fig.~\ref{fig:example1}(b) and Fig.~\ref{fig:example1}(c). It can be clearly observed that the power concentration of the respective target in the angle-delay domain significantly improves, and the equivalent narrowband target model, as represented by \eqref{eqn:2D_swb_conjugated}, is effectively achieved.

It is important to note that the accuracy of the wideband effect compensation relies heavily on the precision of the coarse angle estimate. At higher angles and larger bandwidths, the spread caused by SWE becomes sufficiently large to induce notable coarse angle bin migration. As a result, the rotation-based array processing technique proposed in \cite{wang2018spatial} becomes ineffective under these conditions, which we demonstrate numerically in Section~\ref{sec:results}.

We employ the 2D-OMP-based super-resolution frequency estimation technique as suggested in \cite{rai2025multi} and demonstrate its robustness against angle migration caused by the SWE. We stack the measurements $y_{q}[n]$ for all $1\leq q\leq Q$ in an $Q\times 1$ vector $\mathbf{y}_{n}$ for the $n$-th time sample. Further, define the $Q\times N$ matrix $\mathbf{Y}=[\mathbf{y}_{1},\hdots,\mathbf{y}_{N}]$ and the steering vectors in the range and angular frequency domains as $\mathbf{b}(\Omega_{R})$ and $\mathbf{c}(\Omega_{\theta})$, respectively, as
\par\noindent\small
\begin{align}
    &\mathbf{b}(\Omega_{R})\doteq[1,\exp{(j2\pi\Omega_{R})},\hdots,\exp{(j2\pi\Omega_{R}(N-1))}]^{T},\label{eqn:range steering vector}\\
    &\mathbf{c}(\Omega_{\theta})\doteq[1,\exp{(j2\pi\Omega_{\theta})},\hdots,\exp{(j2\pi\Omega_{\theta}(Q-1))}]^{T}.\label{eqn:angle steering vector}
\end{align}
\normalsize
Then, using \eqref{eqn:2D_swb_conjugated}, we obtain
\par\noindent\small
\begin{align}
    \Tilde{\mathbf{Y}}^k= \widetilde{a}_{k}\mathbf{c}(\bm{\Omega}_{\theta}^k)\mathbf{b}^{T}(\bm{\Omega}_{R}^k)+\mathbf{\Tilde{W}},\label{eqn:range_angle}
\end{align}
\normalsize
Here, $\mathbf{\Tilde{W}}$ represents the $Q\times N$ stacked noise matrix.
Our goal is to recover $\bm{\Omega_{\theta}}$ and $\bm{\Omega_{R}}$ from $\mathbf{\Tilde{Y}}^k$ with super-resolution. To this end, we exploit the sparsity of the target scene. We select grids consisting of $G_{R}$ points $\rho_{1 \leq g \leq G_{R}}$ for potential target range frequencies and $G_{\theta}$ points $\phi_{1 \leq g \leq G_{\theta}}$ for angular frequencies, ensuring both $G_{R}, G_{\theta} \gg 1$ and that discretization errors are negligible. It is important to note that, based on $\Omega_{R} = \gamma \tau_R / f_s$ and $\Omega_{\theta} = d \sin{(\theta)}/\lambda$, the grids $\rho_{1 \leq g \leq G_{R}}$ and $\phi_{1 \leq g \leq G_{\theta}}$ can equivalently represent the potential target ranges and AoAs, respectively. Substituting these grid points in $\mathbf{b}(\cdot)$ and $\mathbf{c}(\cdot)$, we construct the $R\times G_{R}$ sensing matrix $\mathbf{B}=[\mathbf{b}(\rho_{1}),\hdots,\mathbf{b}(\rho_{G_{R}})]$ and the $Q\times G_{\theta}$ sensing matrix $\mathbf{C}=[\mathbf{c}(\phi_{1}),\hdots,\mathbf{c}(\phi_{G_{\theta}})]$. Finally, \eqref{eqn:2D_swb_conjugated} can be expressed as
\par\noindent\small
\begin{align}  \mathbf{\Tilde{Y}^k}=\mathbf{C}\mathbf{Z}\mathbf{B}^{T}+\mathbf{\Tilde{W}},\label{eqn:2D_range_angle}
\end{align}
\normalsize
where the unknown $G_{\theta}\times G_{D}$ matrix $\mathbf{Z}$ contains the coefficients $\{z_{k'}\}$ as well as the target range and AoA information. Specifically, a non-zero element in \(\mathbf{Z}\) indicates the presence of a target, with its AoA and range corresponding to a point in the 2D grid $\{(\phi_{i_{1}}, \rho_{i_{2}}) : 1 \leq i_{1} \leq G_{\theta}, 1 \leq i_{2} \leq G_{R}\}$. Since $ G_{R}G_{\theta} \gg 1$, the matrix $\mathbf{Z}$ is sparse, and hence, the desired range and AoA can be recovered with super-resolution using measurements $\mathbf{Y}$ given matrices $\mathbf{C}$ and $\mathbf{B}$. Note that the joint range-angle estimation has now been reduced to determining the $\textrm{supp}(\mathbf{Z})$. In \eqref{eqn:2D_range_angle}, the measurement matrices \(\mathbf{C}\) and \(\mathbf{B}\), and consequently the recovery performance, are influenced by the selection of grids $\rho_{1 \leq g \leq G_{R}}$ and $\phi_{1 \leq g \leq G_{\theta}}$. Recently, Yu et al. \cite{yu2011measurement} investigated the development of optimal sensing matrices to enhance detection performance in compressive CS-based MIMO radar systems. Building upon this, we employ the low-complexity two-dimensional orthogonal matching pursuit (2D-OMP) algorithm proposed by Fang et al. \cite{fang20122d}, which offers reduced storage requirements and computational complexity compared to the vectorized one-dimensional OMP (1D-OMP) algorithm, making it particularly suitable for large-scale MIMO applications. Once the refined estimates $\hat{\hat{\Omega}}_{R}^k$ and $\hat{\hat{\Omega}}_{\theta}^k$ are obtained, the target's complex coefficient can be computed by a projection-based least squares method as follows
\par\noindent\small
\begin{align}
    \hat{\hat{a}}^k = \mathbf{b}^H(\hat{\hat{\Omega}}_{R}^k)(\mathbf{Y}^w\circ\Theta^*(\hat{\hat{\Omega}}_{\theta}^k))\mathbf{c}^{*}(\hat{\hat{\Omega}}_{\theta}^k)
    \label{eqn:complex_coefficient}
\end{align}
\normalsize
where, $\mathbf{Y}^w$ is the stacked measurement matrix of the spatial wideband signal in \eqref{eqn:2D_swb}, $\mathbf{\Theta}(\Omega_{\theta})\doteq \mathrm{exp}(j2\pi\frac{\alpha}{N}\Omega_{\theta}qn)$ is the wideband phase matrix, $\mathbf{b,c}$ are steering vectors as defined in \eqref{eqn:angle steering vector}, and $\circ$ denotes the element-wise product.

After obtaining the refined signature of the $k$-th path, $\lbrace \hat{\hat{a}}^k, \hat{\hat{\Omega}}_{R}^k, \hat{\hat{\Omega}}_{\theta}^k \rbrace$, we remove the corresponding contribution from the IF signal. This operation ensures that the associated peak is eliminated, allowing the algorithm to detect remaining paths in subsequent iterations. Consequently, we update $\Check{y}_q[n]$ as follows
\par\noindent\small
\begin{align}
    \Check{y}_q[n] = y_q[n]-\hat{\hat{a}}^k \exp({j2\pi \hat{\hat{\Omega}}_{R}^k}n) \exp{(j2\pi \hat{\hat{\Omega}}_{\theta}^k}q) \exp\left({j2\pi \hat{\hat{\Omega}}_{\theta}^k}\frac{\alpha}{N}q\right).
    \label{eqn:eliminate}
\end{align}
\normalsize

We summarize the stepwise procedure in Algorithm\ref{alg:swb}.

\begin{algorithm}
\caption{Coarse-to-Fine Signature Estimation}
\label{alg:swb}
\begin{algorithmic}[1]
\STATE Input: $y_q[n] \hspace{1 mm}\forall n,q, G_R, G_{\theta}$
\STATE Take the 2D-DFT of the antenna-fast time IF signal.
\STATE Find the number of peaks and corresponding range-angle bins to get coarse signature $\lbrace \hat{K},\hat{\Omega}_{R}^k,\hat{\Omega}_{\theta}^k\rbrace$
\STATE Initialize $\Check{y}_q[n] = y_q[n]$
\STATE Construct the overcomplete dictionaries $\mathbf{B}$ and $\mathbf{C}$ using the range-angle steering vectors in \eqref{eqn:angle steering vector}. 
\FOR{$k = 1$ to $\hat{K}$}
    \STATE Remove the SWE using conjugation by
    \eqref{eqn:conjugation} and then select the IF signal corresponding to the $k^{th}$ path.
    \STATE Apply 2D-OMP method with a fixed sparsity level of 1 to estimate the fine-tuned range-angular frequencies $\hat{\hat{\Omega}}_{R}^k,\hat{\hat{\Omega}}_{\theta}^k$ 
    \STATE Find the corresponding target complex coefficient $\hat{\hat{a}}^k$ using \eqref{eqn:complex_coefficient}
    \STATE Eliminate the current path from the observation matrix using \eqref{eqn:eliminate}
\ENDFOR
\STATE Output: $\lbrace \hat{K}, \hat{\hat{a}}^k,\hat{\hat{\Omega}}_{R}^k,\hat{\hat{\Omega}}_{\theta}^k\rbrace$
\end{algorithmic}
\end{algorithm}

\section{Results}
\label{sec:results}

The normalized bandwidth parameter is set to $\alpha = 0.05$, and the carrier frequency is fixed at $f_c = 77 \mathrm{GHz}$. A virtual array consisting of $Q = 64$ elements is considered. Additionally, the number of fast-time samples per chirp is set to $N = 512$, unless otherwise specified. For the chosen XL-MIMO configuration, the corresponding near-field distance is below the minimum simulated target range, ensuring that all targets lie strictly in the far-field regime and validating the planar wavefront assumption used throughout this work. Moreover, numerical exploration reveals that, the two target blocks remain non-overlapping even if they are separated by $(1^{\degree},~0.1~\mathrm{m})$.

\subsection*{Example: target localization}
\begin{figure}
    \centering
    \begin{subfigure}[b]{0.8\linewidth}
        \centering
        \includegraphics[width=0.8\linewidth]{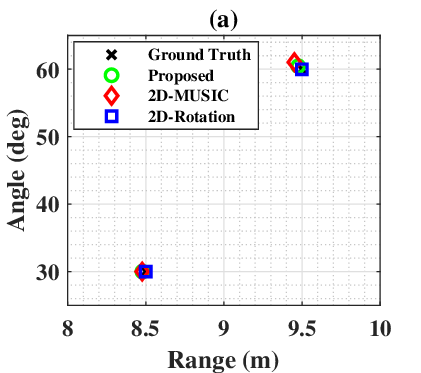} 
    \end{subfigure}
    \hfill
    \begin{subfigure}[b]{0.8\linewidth}
        \centering
        \includegraphics[width=0.8\linewidth]{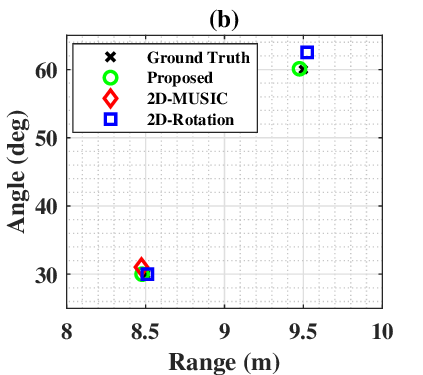}
    \end{subfigure}
    \caption{Target localization at (a) $\alpha = 0.01$ (b) $\alpha = 0.05$.}
    \label{fig:localization}
\end{figure}

We present an example of localizing $K = 2$ targets in Fig.~\ref{fig:localization}. In the spatial narrowband case (Fig.~\ref{fig:localization}(a)), the 2D-MUSIC, 2D-Rotation, and the proposed method all accurately localize the two targets in the range-angle domain. However, under the spatial wideband scenario, the narrowband 2D-MUSIC algorithm detects only one target, while the 2D-Rotation method exhibits a localization error of approximately $2.5^{\circ}$ for the second target, as illustrated in Fig.~\ref{fig:localization}(b). In contrast, the proposed method successfully localizes both targets with high accuracy, placing them in $0.01^{\circ}$ proximity to their actual positions in the spatial wideband case.

\begin{figure*}  
    \centering
    \begin{subfigure}[b]{0.31\linewidth}
        \centering
        \includegraphics[width=\linewidth]{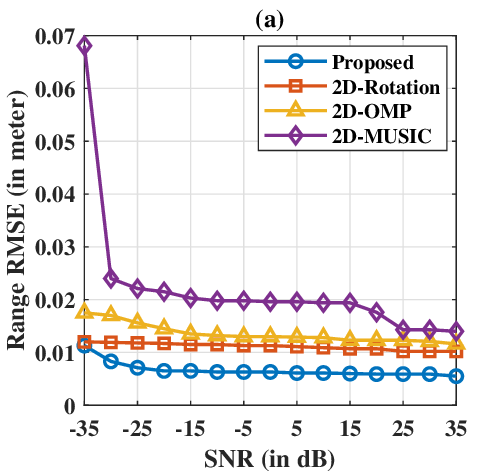} 
    \end{subfigure}
    \hfill
    \begin{subfigure}[b]{0.31\linewidth}
        \centering
        \includegraphics[width=\linewidth]{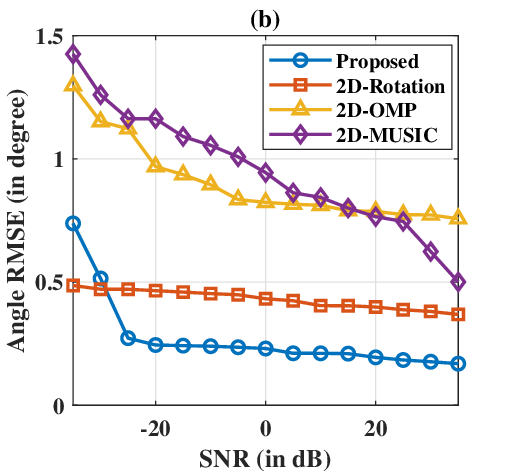}
    \end{subfigure}
    \hfill
    \begin{subfigure}[b]{0.31\linewidth}
        \centering
        \includegraphics[width=\linewidth]{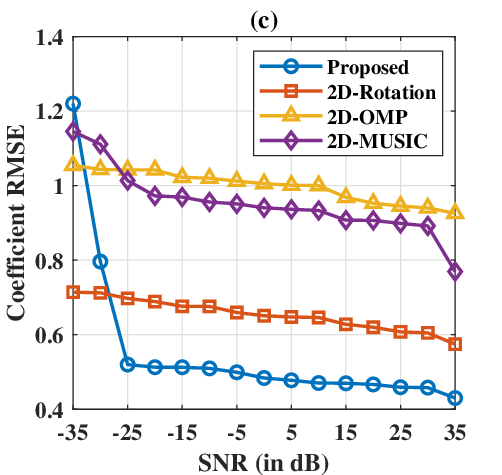}
    \end{subfigure}
    \caption{RMSE of the estimated parameters at $\alpha = 0.05$ (a) range (b) AoA (c) complex coefficient.}
    \label{fig:rmse}
\end{figure*}

\begin{figure}
    \centering
    \begin{subfigure}[b]{0.8\linewidth}
        \centering
        \includegraphics[width=0.8\linewidth]{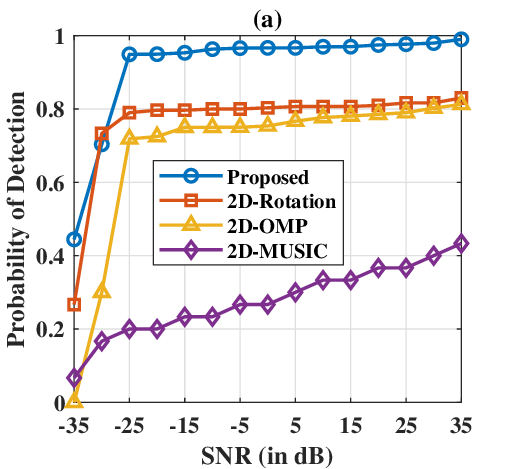} 
    \end{subfigure}
    \hfill
    \begin{subfigure}[b]{0.8\linewidth}
        \centering
        \includegraphics[width=0.8\linewidth]{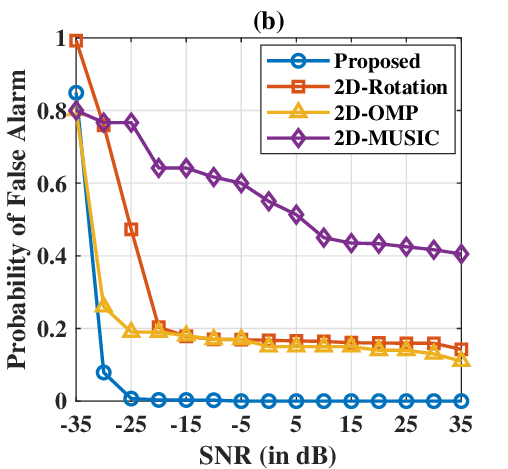}
    \end{subfigure}
    \caption{(a) Probability of detection and (b) probability of false alarm at $\alpha = 0.05$.}
    \label{fig:hit_false_rate}
\end{figure}

\subsection*{Performance analysis: signature estimation}
We evaluate the effectiveness of our proposed method by measuring its probability of detection, probability of false alarm, and root mean square error (RMSE). The target scenario consists of $K = 3$ far-field point targets with ranges uniformly drawn from the interval $[8\,\mathrm{m}, 20\,\mathrm{m}]$ and angles of arrival (AoAs) uniformly drawn from $[10^{\circ}, 80^{\circ}]$. The target complex gains are modeled as $a_{k} = \exp{\left(j\psi_{k}\right)}$, where $\psi_{k} \sim \mathcal{U}[0, 2\pi)$, while the noise samples are drawn as $w_{q}[t] \sim \mathcal{CN}(\mathbf{0}, \sigma^{2}\mathbf{I})$, independently across all virtual array elements and time samples.

We evaluate the performance of the proposed CS-based estimation method for XL-MIMO FMCW radar under the SWE by comparing it with three benchmark techniques: the rotation-based spatial wideband estimation method from \cite{wang2018spatial}, the 2D-MUSIC algorithm \cite{belfiori20122d}, and the conventional 2D-OMP method \cite{author2024}. The comparison is conducted in terms of target detection probability and parameter estimation accuracy.

A detected target is classified as a hit if its estimated range and AoA fall within the predefined resolution limits ($0.02$ m and $1.5^{\circ}$ in this experiment); otherwise, it is considered a false alarm. The probability of detection and false alarm for different methods is presented in Fig. \ref{fig:hit_false_rate}, while the corresponding RMSE values, computed for correctly detected targets, are shown in Fig. \ref{fig:rmse}.

As shown in Fig.~\ref{fig:rmse}(a), the proposed method achieves substantially lower range RMSE across the entire SNR range, with rapid error reduction beyond $-15$\,dB, while 2D-MUSIC and 2D-OMP exhibit consistently higher errors due to their inability to handle the SWE-induced coupling. The 2D-Rotation method improves upon these narrowband techniques but remains inferior to the proposed approach.

A similar trend is observed in Fig.~\ref{fig:rmse}(b) for AoA estimation, where the proposed method achieves sub-$0.3^{\circ}$ RMSE at high SNRs, while 2D-MUSIC and 2D-OMP perform poorly across all SNRs. The 2D-Rotation method offers moderate improvements but fails to match the accuracy of the proposed method.

In Fig.~\ref{fig:rmse}(c), the RMSE of the estimated complex reflection coefficients is plotted against SNR. Once again, the proposed method exhibits superior performance, with significant error reduction at higher SNR values, while the other three techniques maintain relatively high and flat RMSE profiles. These results collectively demonstrate the robustness and effectiveness of the proposed method for accurate target signature estimation under spatial wideband effects in XL-MIMO FMCW radar systems.

Fig.~\ref{fig:hit_false_rate} illustrates the target detection performance in terms of probability of detection and false alarm versus SNR for $\alpha = 0.0.5$. As shown in Fig.~\ref{fig:hit_false_rate}(a), the proposed method achieves a probability of detection close to 1 at SNRs above $-10$\,dB, with a rapid performance improvement as SNR increases. In contrast, the 2D-Rotation method maintains a moderate probability of detection of approximately 70–80\% across the entire SNR range, while the 2D-OMP method performs inferior to both the proposed and the 2D-Rotation method. Furthermore, the 2D-MUSIC consistently fails to reliably detect targets under spatial wideband conditions, achieving a probability of detection below 50\%.

In terms of false detection, depicted in Fig.~\ref{fig:hit_false_rate}(b), the proposed method exhibits a sharp reduction in probability of false alarm with increasing SNR, achieving near-zero false alarms beyond -15\,dB. The 2D-Rotation method and 2D-OMP, however, retains a high probability of false alarm of around 20\%, while 2D-MUSIC show persistently high probability of false alarm (around 40–60\%) across all SNR values. These results clearly demonstrate the superiority of the proposed compressive sensing-based estimation technique for SWE-affected XL-MIMO FMCW radar, delivering both high detection reliability and low probability of false alarm, even under challenging spatial wideband conditions.

\subsection*{Complexity: run-time}

Using the 2D-MUSIC algorithm for 2D range-angle parameter estimation with $Q = 64$, $N = 512$, and $K = 3$ takes approximately 54 seconds on a 10\textsuperscript{th}-generation Intel i7 processor. In comparison, the 2D-Rotation method performs the same task in just 0.57 seconds. Both of these methods, however, are limited to spatial narrowband scenarios. The proposed method not only reduces the estimation time to 0.17 seconds but also extends applicability to spatial wideband cases, offering both improved efficiency and broader utility.

\section{Conclusion} 
In this work, we have proposed a low-complexity joint range-angle estimation methodology for XL-MIMO FMCW radars under the SWE, capable of simultaneously estimating the number of targets and their complex reflection coefficients. Numerical results demonstrate that the proposed technique achieves a significantly higher probability of detection and a lower probability of false alarm rate, along with lower MSE, compared to existing methods. Moreover, the scalability and computational efficiency of the proposed approach make it well-suited for a range of localization-based applications across both low and high-bandwidth settings. Additionally, the proposed algorithm is computationally efficient, as it does not require any supplementary mechanisms to mitigate range-angle bin migration effects induced by SWE.

Future work will focus on extending the proposed framework to jointly address the spatial wideband and near-field effects, enabling accurate parameter estimation for the XL-MIMO radar operating in short-range regimes. Additionally, the current assumption of non-overlapping SWE-induced target blocks will be relaxed by developing robust algorithms capable of resolving overlapping scatter responses.


\bibliographystyle{IEEEtran}
\bibliography{references}

\begin{thebibliography}{10}
\providecommand{\url}[1]{#1}
\csname url@samestyle\endcsname
\providecommand{\newblock}{\relax}
\providecommand{\bibinfo}[2]{#2}
\providecommand{\BIBentrySTDinterwordspacing}{\spaceskip=0pt\relax}
\providecommand{\BIBentryALTinterwordstretchfactor}{4}
\providecommand{\BIBentryALTinterwordspacing}{\spaceskip=\fontdimen2\font plus
\BIBentryALTinterwordstretchfactor\fontdimen3\font minus
  \fontdimen4\font\relax}
\providecommand{\BIBforeignlanguage}[2]{{%
\expandafter\ifx\csname l@#1\endcsname\relax
\typeout{** WARNING: IEEEtran.bst: No hyphenation pattern has been}%
\typeout{** loaded for the language `#1'. Using the pattern for}%
\typeout{** the default language instead.}%
\else
\language=\csname l@#1\endcsname
\fi
#2}}
\providecommand{\BIBdecl}{\relax}
\BIBdecl

\bibitem{patole2017automotive}
S.~M. Patole, M.~Torlak, D.~Wang, and M.~Ali, ``Automotive radars: A review of
  signal processing techniques,'' \emph{IEEE Signal Processing Magazine},
  vol.~34, no.~2, pp. 22--35, 2017.

\bibitem{wu2021intelligent}
Q.~Wu, S.~Zhang, B.~Zheng, C.~You, and R.~Zhang, ``Intelligent reflecting
  surface-aided wireless communications: A tutorial,'' \emph{IEEE transactions
  on communications}, vol.~69, no.~5, pp. 3313--3351, 2021.

\bibitem{park2024spatial}
J.-H. Park, S.~Lee, G.~Moon, and S.-C. Kim, ``Spatial-wideband effect
  compensation for high resolution imaging in mimo fmcw radar,'' \emph{IEEE
  Transactions on Instrumentation and Measurement}, 2024.

\bibitem{han2023range}
K.~Han and S.~Hong, ``Range-angle decoupling technique using
  wavelength-dependent beamforming for high-resolution mimo radar,'' \emph{IEEE
  Transactions on Microwave Theory and Techniques}, 2023.

\bibitem{xi2021joint}
R.~Xi, C.~Zheng, T.~Huang, L.~Wang, and Y.~Liu, ``Joint range and angle
  estimation for wideband forward-looking imaging radar,'' \emph{IEEE Sensors
  Journal}, vol.~22, no.~1, pp. 446--460, 2021.

\bibitem{durr2020range}
A.~D{\"u}rr, B.~Schneele, D.~Schwarz, and C.~Waldschmidt, ``Range-angle
  coupling and near-field effects of very large arrays in mm-wave imaging
  radars,'' \emph{IEEE Transactions on Microwave Theory and Techniques},
  vol.~69, no.~1, pp. 262--270, 2020.

\bibitem{hu2023range}
Y.~Hu, W.~Deng, Y.~Dong, and X.~Wu, ``Range-angle coupling in linear sparse
  array: Far-field model with narrow-band fmcw signal,'' \emph{IEEE
  Transactions on Aerospace and Electronic Systems}, 2023.

\bibitem{wang2021direction}
F.~Wang, Z.~Tian, G.~Leus, and J.~Fang, ``Direction of arrival estimation of
  wideband sources using sparse linear arrays,'' \emph{IEEE Transactions on
  Signal Processing}, vol.~69, pp. 4444--4457, 2021.

\bibitem{wu2024coffee}
X.~Wu, Z.~Yang, Z.~Wei, R.~Schober, and Z.~Xu, ``Coffee: Covariance fitting and
  focusing for wideband direction-of-arrival estimation,'' \emph{IEEE
  Transactions on Signal Processing}, 2024.

\bibitem{janoudi2023signal}
V.~Janoudi, P.~Schoeder, T.~Grebner, N.~Appenrodt, J.~Dickmann, and
  C.~Waldschmidt, ``Signal model for coherent processing of uncoupled and low
  frequency coupled mimo radar networks,'' \emph{IEEE Journal of Microwaves},
  vol.~4, no.~1, pp. 69--85, 2023.

\bibitem{wang2018spatial}
B.~Wang, F.~Gao, S.~Jin, H.~Lin, and G.~Y. Li, ``Spatial-and frequency-wideband
  effects in millimeter-wave massive mimo systems,'' \emph{IEEE Transactions on
  Signal Processing}, vol.~66, no.~13, pp. 3393--3406, 2018.

\bibitem{wei2021channel}
X.~Wei and L.~Dai, ``Channel estimation for extremely large-scale massive mimo:
  Far-field, near-field, or hybrid-field?'' \emph{IEEE Communications Letters},
  vol.~26, no.~1, pp. 177--181, 2021.

\bibitem{fang2019joint}
W.-H. Fang and L.-D. Fang, ``Joint angle and range estimation with signal
  clustering in fmcw radar,'' \emph{IEEE Sensors Journal}, vol.~20, no.~4, pp.
  1882--1892, 2019.

\bibitem{wang1985coherent}
H.~Wang and M.~Kaveh, ``Coherent signal-subspace processing for the detection
  and estimation of angles of arrival of multiple wide-band sources,''
  \emph{IEEE Transactions on Acoustics, Speech, and Signal Processing},
  vol.~33, no.~4, pp. 823--831, 1985.

\bibitem{friedlander1993direction}
B.~Friedlander and A.~J. Weiss, ``Direction finding for wide-band signals using
  an interpolated array,'' \emph{IEEE Transactions on Signal Processing},
  vol.~41, no.~4, pp. 1618--1634, 1993.

\bibitem{yoon2006tops}
Y.-S. Yoon, L.~M. Kaplan, and J.~H. McClellan, ``Tops: New doa estimator for
  wideband signals,'' \emph{IEEE Transactions on Signal processing}, vol.~54,
  no.~6, pp. 1977--1989, 2006.

\bibitem{li1997angle}
J.~Li, D.~Zheng, and P.~Stoica, ``Angle and waveform estimation via relax,''
  \emph{IEEE transactions on aerospace and electronic systems}, vol.~33, no.~3,
  pp. 1077--1087, 1997.

\bibitem{lovescu2020fundamentals}
C.~Lovescu and S.~Rao, ``The fundamentals of millimeter wave radar sensors,''
  \emph{Texas Instruments, Julio}, 2020.

\bibitem{rai2025low}
C.~Rai and D.~Sen, ``Low complexity doa-toa signature estimation for
  multi-antenna multi-carrier systems,'' in \emph{ICASSP 2025-2025 IEEE
  International Conference on Acoustics, Speech and Signal Processing
  (ICASSP)}.\hskip 1em plus 0.5em minus 0.4em\relax IEEE, 2025, pp. 1--5.

\bibitem{rai2025multi}
C.~Rai, H.~Singh, and A.~Chattopadhyay, ``Multi-target range, doppler and angle
  estimation in mimo-fmcw radar with limited measurements,'' \emph{IEEE
  Transactions on Signal Processing}, 2025.

\bibitem{rai2025artificial}
C.~Rai and D.~Sen, ``An artificial intelligence enabled signature estimation of
  dual wideband systems in ultra-low signal-to-noise ratio,'' \emph{arXiv
  preprint arXiv:2504.14226}, 2025.

\bibitem{bevelander2025divide}
A.~Bevelander, K.~Batselier, and N.~J. Myers, ``A divide-and-conquer approach
  for sparse recovery in high dimensions,'' in \emph{ICASSP 2025-2025 IEEE
  International Conference on Acoustics, Speech and Signal Processing
  (ICASSP)}.\hskip 1em plus 0.5em minus 0.4em\relax IEEE, 2025, pp. 1--5.

\bibitem{MATLABCode}
K.~Tikuišis, ``peaks2 - find peaks in 2d data without additional toolbox,''
  MATLAB code, 2023, available at
  \url{https://www.mathworks.com/matlabcentral/fileexchange/113225-peaks2-find-peaks-in-2d-data-without-additional-toolbox}.

\bibitem{rai2025two}
C.~Rai and D.~Sen, ``A two-stage rotation-based super-resolution signature
  estimation for spatial wideband systems,'' \emph{arXiv preprint
  arXiv:2503.18111}, 2025.

\bibitem{yu2011measurement}
Y.~Yu, A.~P. Petropulu, and H.~V. Poor, ``Measurement matrix design for
  compressive sensing--based mimo radar,'' \emph{IEEE Transactions on Signal
  Processing}, vol.~59, no.~11, pp. 5338--5352, 2011.

\bibitem{fang20122d}
Y.~Fang, J.~Wu, and B.~Huang, ``2d sparse signal recovery via 2d orthogonal
  matching pursuit,'' \emph{Science China Information Sciences}, vol.~55, pp.
  889--897, 2012.

\bibitem{belfiori20122d}
F.~Belfiori, W.~van Rossum, and P.~Hoogeboom, ``2d-music technique applied to a
  coherent fmcw mimo radar,'' in \emph{IET International Conference on Radar
  Systems (Radar 2012)}.\hskip 1em plus 0.5em minus 0.4em\relax IET, 2012, pp.
  1--6.

\bibitem{author2024}
C.~Rai and D.~Sen, ``Low-complexity doa-toa signature estimation for
  multi-antenna multi-carrier systems,'' to appear in ICASSP, 2025.

\end{thebibliography}

\end{document}